\documentclass[preprint2]{aastex}

\slugcomment{Not to appear in Nonlearned J., 45.}

\shorttitle{GWs from DCOs} \shortauthors{Djorgovski et al.}

\begin{document}


\title{Gravitational-wave radiation from double compact objects with eLISA in the Galaxy.}


\author{Jinzhong Liu\altaffilmark{1} and Yu Zhang\altaffilmark{1}}
\affil{}

\altaffiltext{1}{National Astronomical Observatories/Xinjiang
Observatory, Chinese Academy of Sciences, 150 Science 1--street
Urumqi, 830011 Xinjiang, P. R. China. liujinzh@xao.ac.cn}
% \altaffiltext{2}{Society of
%Fellows, Harvard University.} \altaffiltext{3}{present address:
%Center for Astrophysics,
 %   60 Garden Street, Cambridge, MA 02138}
%\altaffiltext{4}{Visiting Programmer, Space Telescope Science Institute}
%\altaffiltext{5}{Patron, Alonso's Bar and Grill}


\begin{abstract}
The phase of in-spiral of double compact objects (DCOs: NS+WD,
NS+NS, BH+NS, and BH+BH binaries) in the disk field population of
the Galaxy provides a potential source in the frequency range from
$10^{-4}$ to 0.1 Hz, which can be detected by the European New
Gravitational Observatory (NGO: eLISA is derived from the previous
LISA proposal) project. In this frequency range, much stronger
gravitational wave (GW) radiation can be obtained from DCO sources
because they possess more mass than other compact binaries (e.g.,
close double white dwarfs). In this study, we aim to calculate the
gravitational wave signals from the resolvable DCO sources in the
Galaxy using a binary population synthesis approach, and to carry
out physical properties of these binaries using Monte Carlo
simulations. Combining the sensitivity curve of the eLISA detector
and a confusion-limited noise floor of close double white dwarfs, we
find that only a handful of DCO sources can be detected by the eLISA
detector. The detectable number of DCO sources reaches 160, in the
context of low-frequency eLISA observations we find that the number
of NS+WD, NS+NS, BH+NS, and BH+BH are 132, 16, 3, and 6,
respectively.
\end{abstract}





\section{Introduction}

Gravitational waves (GWs) are a natural result of Einstein's theory
of gravity resulting from a space perturbation of the metric
traveling at the speed of light. This phenomenon of space-time has
still not been directly observed on the ground such as LIGO or
VIRGO, because there exist seismic and gravity gradient noise. The
observation of the binary pulsar PSR 1913+16, which is a neutron
star plus neutron star (NS+NS) system, has given an indirect
evidence of GW radiation \citep{hulse75,tay82}. Therefore, the
European New Gravitational Wave Observatory (NGO is referred to as
eLISA) mission will search the GW radiation in the frequency band
between $10^{-4}$ Hz and 0.1 Hz, which is the previous LISA's
heritage ({Amaro-Seoane et al 2012, and references therein). The
main GW sources in this frequency range are extreme mass ratio
inspirals (EMRIs) of stellar-mass compact objects orbiting the
massive black holes (BHs) \citep[e.g.,][]{glam2005, hopm2006}, the
coalescence of super--massive BH binaries of merging galaxies
\citep[e.g.,][]{iwas2011, liu2012} and Galactic compact double
binaries \citep[e.g.,][]{hils1990, ruit2010}. For example, due to
the largest population in the Galaxy, close double white dwarfs
(DWDs) are believed to dominate the Galactic GW foreground radiation
that generate a confusion--limited noise floor for the classical
LISA detector, with several thousand of the higher GW radiation
signal sources being possibly resolved \citep[e.g.,][]{evan1987,
hils1990, nele2001, liu2009, liu2010, liua2010,litt2011,niss2012,
shah2012, shah2013, nelemans2013}. In this study, we focus on other
types of double compact binaries (DCOs): neutron star plus white
dwarf (NS+WD), double neutron star binaries (NS+NS), black hole plus
neutron star (BH+NS), and double black hole (BH+BH) binaries. Few
studies investigate the importance of these GW sources because these
objects are much rarer than the DWD binaries. However, DCOs can
radiate stronger GW signals because of higher mass than WDs.

 As a physical reality event,
DCOs play an important role in stellar evolution of population
synthesis studies \citep{tutu1994, dewi2003, wang2009, chen2011,
zhan2012, jian2012}, because they are expected to be potential
progenitors of related objects, such as ultra--compact X--ray
binaries \citep[e.g.,][]{vand2005} and short--hard $\gamma$--ray
bursts \citep[e.g.,][]{eich1989,pacz1991,nara1992}. Meanwhile, the
final merger processes of the DCOs are expected to be a type of high
frequency GW sources for the ground-based GW detectors (such as LIGO
or VIRGO). In this class of compact binaries, their merger rates and
birth rates are still an open question, especially for systems
containing BHs \citep{abadi2010}. Several double NSs (NS+NS) are
currently detected through searching binary pulsars, only PRS
J0703--3039 \citep{burg2003} belongs to the eLISA frequency ranges,
and BH or BH+NS systems have not been observed so far. Therefore a
binary population synthesis (BPS) approach has been applied to
evaluate the merger rates of DCOs \citep{hurl2000, hurl2002,
han2007}, and this method has also been systematically used to study
the GW radiation in the Galaxy \citep{nele2001, liu2009}.

In this study, we analyze the DCO systems as GW sources in the disk
field population that obtained from a BPS model, and address the GW
radiation from these sources comparing with the sensitivity curve of
the eLISA detector and the confusion--limited noise floor of DWD
systems. This study can be used to determine the binary parameters
 when the BH/NS in--spiral processes can be detected
by GW detectors in future. In the next section, the model is
described. In Sect. 3, we present the results and discussion in our
simulations.

\section{Model}

The stellar evolution was based on the work previously presented by
\citet{hurl2000} and \citet{hurl2002}. The method of calculating GW
radiation from double white dwarf systems,
   combined with the noise curve of LISA, was described by \citet{liu2009} and \citet{liu2010}. Here we use this BPS method \citep{liu2009, liu2010} in the same way to investigate
   the importance of GW radiation from DOC systems with the eLISA in the Galaxy. Below we briefly summarise this method in the current study.

In the BPS model, the rapid binary stellar evolution (BSE) code
\citep{hurl2000, hurl2002} was used to investigate the evolution of
binary stars. This BSE code was built on the Cambridge stellar
evolution tracks \citep{eggl1971, eggl1972, eggl1973}, and the input
physics were updated by \citet{han1994} and \citet{pols1995}, and
various initial distributions of components are performed using a
Monte Carlo simulation simulation including the initial mass
function (IMF) of primary, the initial mass ratio, the initial
orbital separation, the initial eccentricity and the binary space
model in the Galaxy. This code provides an opportunity for
calculating the evolution of a binary by given its zero--age
main--sequence mass (ZAMS) and metallicity. For these DCO sources,
we need to obtain the evolutionary results at the formation of a
stellar remnant: a WD, a NS, or a BH. Some of the most relevant
features (e.g. wind accretion, orbital changes due to mass
variations, tidal evolution, magnetic braking, supernovae kicks,
roche lobe overflow, common-envelope evolution, coalescence of
common-envelope cores, and collision outcomes) of the BSE code can
be found in \citet{hurl2002}.

For the component WD of NS+WD system, three types of WDs are
distinguished in the BSE simulation: a He WD (formed by complete
envelope loss of a first giant branch star with mass less than the
maximum initial mass of igniting in a helium flash), a CO WD (formed
by envelope loss of a thermally pulsing asymptotic giant branch
star), and Oxygen/Neon WD (envelope loss of a thermally pulsing
asymptotic giant branch star). Note that if the CO WD accretes
enough mass $M_{\rm acc}$, then this CO WD will explodes without
leaving a remnant, wile an Oxygen/Neon WD leaves a NS depending on
the hypotheses on the use of accretion-induced collapse (AIC)
(Nomoto \& Kondo 1991). In the BSE code, the valve of Chandrasekhar
is always given by $M_{\rm Ch}=1.44 M_{\odot}$.

 If a NS or BH is obtained from the BPS simulation, its
gravitational mass can be evaluated by
\begin{equation}
M_{\rm NS}=1.17+0.09M_{\rm c,SN},
\end{equation}
where $M_{\rm c,SN}$ is the mass of the CO-core when the supernova
explosion is occurred.

During  an asymmetry explosion process, a velocity kick can be
produced from an explosion with leaving a remnant (a NS or a BH)
\citep{lyne1994}. To obtain the kick velocity
$\textbf{\emph{v}}_{\rm k}$, we choose the kick speed from a
Maxwellian distribution
\begin{equation}
P(v_{k})=\sqrt{\frac{2}{\pi}}\frac{v_{k}^{2}}{\sigma_{k}^{3}}e^{-v_{k}^{2}/2\sigma_{k}^{2}}.
\end{equation}
We use velocity dispersion $\sigma_{k}=$190km/s, which is consistent
with the data on pulsar proper motions \citep{hans1997}. Note that
we only refer to the DCO objects in the Galaxy field population
using a thin disk model (Equ. 13 of Liu 2009), and exclude the DCOs
resided in halo and bulge from our simulation. That is because that
most of these DCO binaries have long orbital periods ($P_{\rm orb}
> 5.6$\,hrs corresponding to log $f <- 4.0$\,Hz) leading to
relative small contributions to GW signals \citep{belc2010}. And the
dynamical interacting (e.g., evolution of stars in the globular
clusters) in the Galaxy is not considered in the current study. We
note that the binary stellar evolutionary channels lead to the
production of DCOs with circular orbits, even if the ZAMS
eccentricity is non--zero in the simulation, because tidal
circularization and synchronization are rapid when a system contains
a near-Roche lobe-filling convective star. This is an assumption in
the BSE code.

\vspace{2mm} \noindent {\footnotesize{\bf Table 1}\quad Parameters
for three types of compact remnant objects obtained from the binary
evolution.
}\vspace{-5mm}\\
\begin{center}
  \footnotesize %\doublerulesep 0.2pt \tabcolsep 23pt
\begin{tabular}{ccclclc}
\hline\hline
Initial mass  &remnant  & mean remnant mass  \\
$[M_{\odot}]$&object &$[M_{\odot}]$\\
\hline
0.32$<M<$ 8--12     &WD            &0.58            \\
  8--12$<M<$ 25--45 &NS    &1.39         \\
25--45$<M<$ 100 &BH    &9.5\\
\hline\hline
\end{tabular}
\end{center}

By means of a BSE code and a Monte Carlo simulation, we get three
types of remnant objects, that is, WD, NS, and BH. Meanwhile, the
initial basic parameters and descriptions are as follows in the BSE
code: the tidal enhancement of the stellar wind \emph{B} is 1000,
the mass transfer efficiency for stable Roche lobe overflow (RLOF)
is 0.5, the ejection efficiency parameter of a common envelope (CE)
is 1, the stellar wind velocity is 20km/s, and the solar metallicity
Z is 0.02. And we also assume a constant star formation rate (SFR)
over the last 13.7\,Gyr in the simulation. The detailed descriptions
of these parameters can be found in some studies \citep{han1998,
hurl2002}. Therefore, we can trace the evolution of these objects
using the BSE code and calculate the boundaries for the initial
masses of progenitors of WDs, NSs and BHs. The mass parameters for
these three types of DCO objects are summarized in Table 1. Indeed,
these acceptable vales of initial mass and mean remnant mass listed
in Table 1 are in common with other studies \citep{han1998,
ziol2010}. In addition, Belczynski et al. (2010) have investigated
the importance of supernova kicks in BH or NS formation processes.
We assume $\sigma_{k}=\,$190km/s as a input parameter in the BSE
code.

In this study, a two year mission lifetime is assumed, a width of a
resolvable frequency bin is $\Delta f$ = 1/$T_{\rm obs}=1.6\times
10^{-8}$\,Hz.
%And based on the sensitivity curve of eLISA detector
%(Petiteau et al. 2008), which averaged over all sky locations and
%polarizations and did not include the foreground noise of close
%white dwarfs,
Petiteau et al.( 2008) showed the eLISA/NGO sensitivity curve, which
averaged over all sky locations and polarizations and did not
include the foreground noise of close white dwarfs. And based on
this sensitivity curve, we creat the time series for the source
signals and add the individual time series of the calculated sources
to produce the total data stream. Similar calculations of the GW
signal analysis combined with a GW detector can be found in Timpano
et al. (2006), liu (2009) and Nissanke et al. (2012). Note that we
should select all DCO systems that have a signal above the
(signal--to--noise ratio: S/N\,=\,7) sensitivity limit of the eLISA
detector. In this study, S/N$>$7 is considered the lowest acceptable
threshold.
\section{Results and discussion}

 Basing on a population
synthesis code, from a sample of $10^{7}$ binaries, we obtain DCO
(NS+WD, NS+NS, BH+NS, and BH+BH) systems. According to the
evolutionary trace of these four types of DCOs in the BSE code, we
give the descriptions of physical properties (e.g., birth rates,
distributions of orbital frequency, and chirp mass) and the GW
radiation contributions of these DCO sources.

 On the basis of evidence in section 2,
in table 2 we list the Galactic birth rates and total numbers for
four types of DCOs from the BPS simulation and the number of systems
that can be detected by the eLISA detector. From this, we see that
the detectable number of all DCO sources is 157. The NS+WD
detectable sources reach 84.1\% of the total DCOs, the percentages
are 10.2\% for NS+NS systems, and 5.70\% for BH+NS and BH+BH
binaries. Within the entire DCO binaries predicted for the present
time in our Galaxy disk,
 NS+WD sources are supreme (73.7\%), with a chief contribution of NS+NS (25.8\%) systems, and a very small
 fractions
 of BH+NS and BH+BH (0.500\%) sources. We note that investigating the different input parameters from BPS simulation can
 change the maximum percentage uncertainty of DCO systems up to a factor of $\sim3.25$. This shows that the number of DCO systems is consistent with
 the Galactic birth rate in table 2. Note that the number quoted here is based on the assumption that one binary with
 $M_{1}>0.8M_{\odot}$ is formed per year in our Galaxy \citep{iben84, yung93}, and one assumes a constant star formation rate SFR$ = 5M_{\odot} /\rm yr$,
 where $5M_{\odot}$ stands for the average binary mass. In table 2 the DCO systems' birthrate in the Galaxy is the convolution of the distribution of the delay times
 (DDT) with the star formation rate (SFR) \citep{greg08}. Due to a
constant SFR, the birthrate of DCO systems is only related to the
DDT, which reveals a function relation in different formation times
of DCOs . Finally, in our simulation we also assume a solar
metallicity Z = 0.02 and 100\% binaries. Additionally, the DCO
systems with mass-transfer or merged are excluded from the BPS
simulation.

\vspace{2mm} \noindent {\footnotesize{\bf Table 2}\quad Galactic
birth rates and total numbers for four types of DCOs from the BPS
simulation and the number of systems than can be detected by the
eLISA detector.
}\vspace{-4.7mm}\\
\begin{center}
  \footnotesize %\doublerulesep 0.2pt \tabcolsep 23pt
\begin{tabular}{cccclc}
\hline\hline
Type &  Total &Birth Rate &Detectable sources  \\
&DCOs&($\rm 10^{-5}yr^{-1}$)&above noise \\
 \hline
$[\rm NS, WD]$     &191343            &53 &132          \\
$[\rm NS, NS]$ &66816    &2.8       &16  \\
$[\rm BH, NS]$ &367    &1.3$\times10^{-2}$&3\\
$[\rm BH, BH]$ &876    &0.17&6\\
\hline\hline
\end{tabular}
\end{center}

\begin{figure*}%multicols\BB\B7\BE\B3\CF²\BB\C4ܸ\A1\B6\AF\A3\AC\BBᵼ\D6\C2ͼ\D0λ\F2\D5߱\ED\B8\F1\B6\AAʧ\A1\A3
%ֻ\C4ܵ\B1ǰλ\D6ã\A8[]\D6в\CE\CA\FD\B1\D8\D0\EB\CAǴ\F3дH\A3\A9\A3\AC\D2\F2\B4\CB\D0\E8Ҫ\CAֶ\AF\B5\F7λ\D6\C3

\centering
\includegraphics[height=2.2in,width=3.0in]{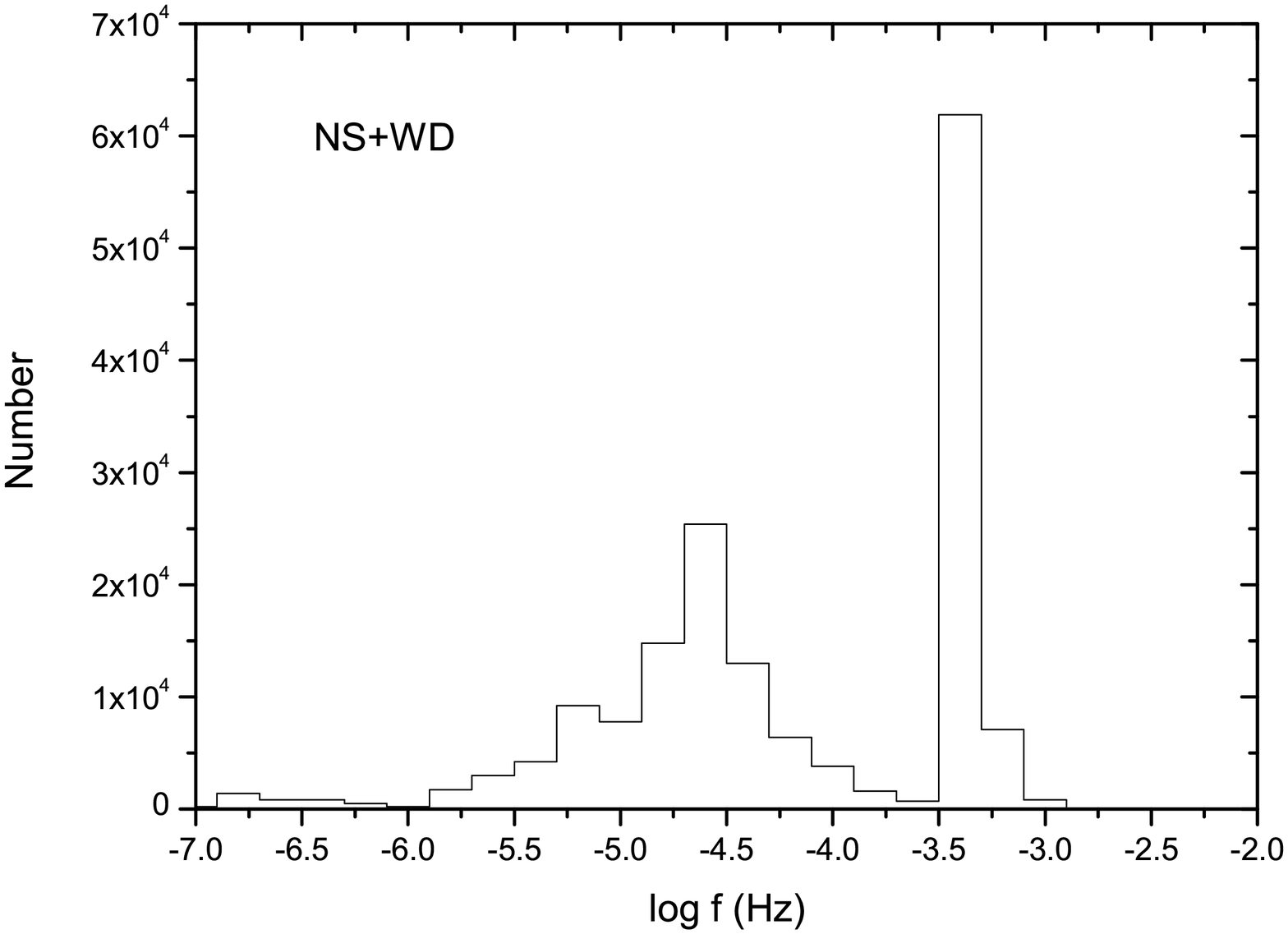}
\includegraphics[height=2.2in,width=3.0in]{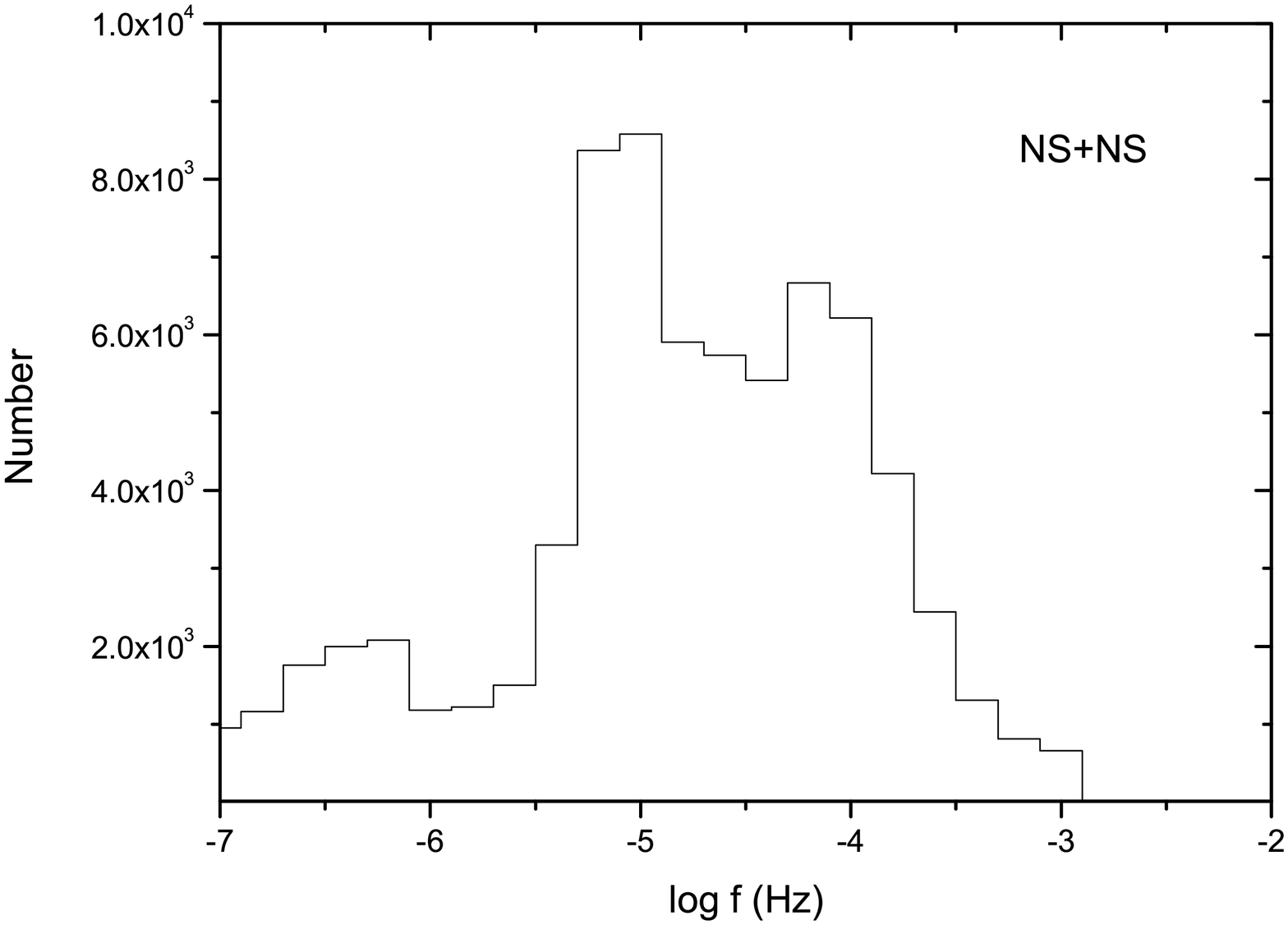}
\includegraphics[height=2.2in,width=3.0in]{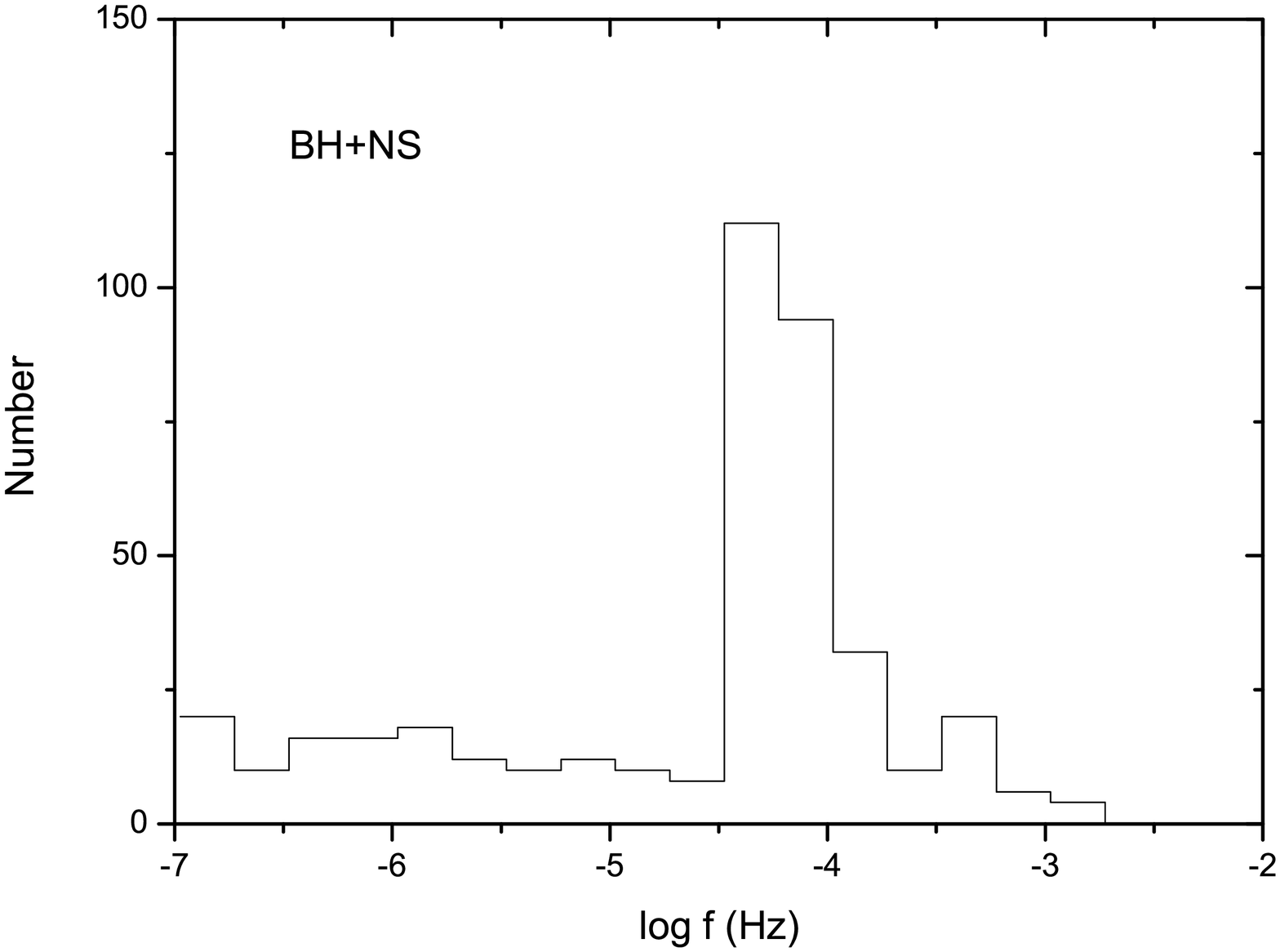}
\includegraphics[height=2.2in,width=3.0in]{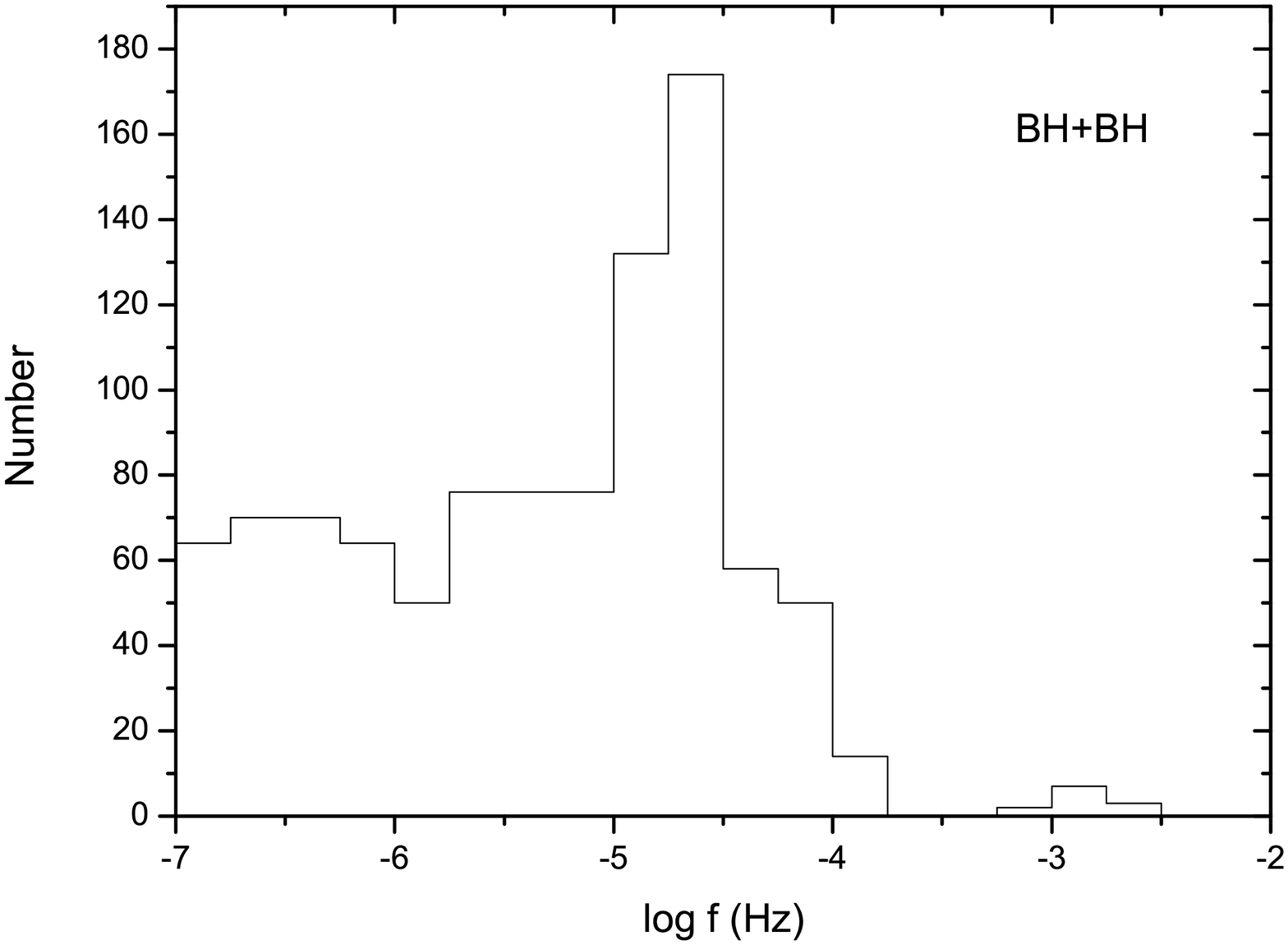}
\caption{Distribution of orbital frequency for four types of DCOs at
the present time in the Galaxy including NS+WD, NS+NS, BH+NS, and
BH+BH.} \label{fig:example2}
\end{figure*}

\subsection{The orbital frequency distribution of DCO systems}
The orbital frequency of a DCO will increase due to GW radiation, so
the orbital frequency is an important physical characteristic in the
present study. In Figure 1, we present the distribution of orbital
frequency for four types of DCOs including NS+WD, NS+NS, BH+NS, and
BH+BH. From this, we see that the orbital frequency distributions
are different for four types of DCO systems, which are characterized
by one or two distinct peaks. For NS+WD systems, a notably higher
narrow peak occurs at log $f =- 3.35$Hz ($P_{\rm orb} \sim
1.24$\,hrs) and nearly 50\% of numbers are less than a vale of
orbital frequency log $f =- 4$\,Hz with a shorter broad peak at log
$f =- 4.6$Hz ($P_{\rm orb} \sim $22.1\,hrs). For NS+NS systems, most
of samples are gathered in the orbital frequency range $10^{-4}$ to
$10^{-5}$\,Hz ($P_{\rm orb} \sim 5.6-55.6$\,hrs). For BH+NS systems,
a high peak is appeared in the orbital frequency log $f =- 4.2$\,Hz
($P_{\rm orb} \sim $8.8hrs). For BH+BH systems, we note that the
orbital frequency distribution is similar to the case of BH+NS, the
differences are not very large besides the peak value log $f =-
4.7$\,Hz ($P_{\rm orb} \sim $27.8\,hrs).

We can know two kinds of information from Figure 1. Firstly,
\citet{pods2003} shown that the formation of a NS or a BH in a
binary system should experience a CE event, and this time-scale in
formation of CE process is often dynamically unstable
\citep{pacz1976}.
 If double NS (or BH) binaries that enter the RLOF as events,
they always coalesce immediately as possible $\gamma$-ray bursts. In
this present work, we assume that if there is not enough orbital
energy to release the CE, it leads to a merger when a common
envelope is formed. We note that the formation channel for a maximum
value in each panel of Figure 1 is shaped by the CE phase.
Specially, at orbital frequencies above $10^{-3}$Hz (corresponding
to a shorter orbital period less than 0.56 h), the DCO systems are
obtained from two successive CE evolutionary phase, and they are
resolved sources (See section 3.3). This is because that dynamically
unstable mass transfer is expected to lead to the formation of a CE.
And because of more CE evolution phases, the evolution of more
massive progenitors trend to produce DCO binaries with shorter
orbital periods \citep{han1998}. Secondly, a synchronization
time-scale \citep{hurl2002} can affect the formation of DCO
binaries, because the components may be spun up by mass transfer.
For example, a WD-NS binary with a separation 10 times the WD radius
would have a very long circularization time-scale ( $\sim 10
^{14}$yr), thus the degenerate damping is dominant only for WD-NS
systems in which the separation can become very small. For the
extreme conditions of short orbital period sources (e.g, AM CVn
stars and ultra-compact X-ray binaries) in the BPS simulations, we
note that the information on the physics of tides and the stability
of the mass-transfer processes is not investigated in the current
study.

%Thirdly, if a binary survives the supernova explosion, then it is
%quite likely that the orbital parameters (e.g., the separation of
%the components), obtain extremely from those of the initial
%conditions.

\begin{figure*}
\centering
\includegraphics[bb=250 100 800 600,height=2.2in,width=2.3in]{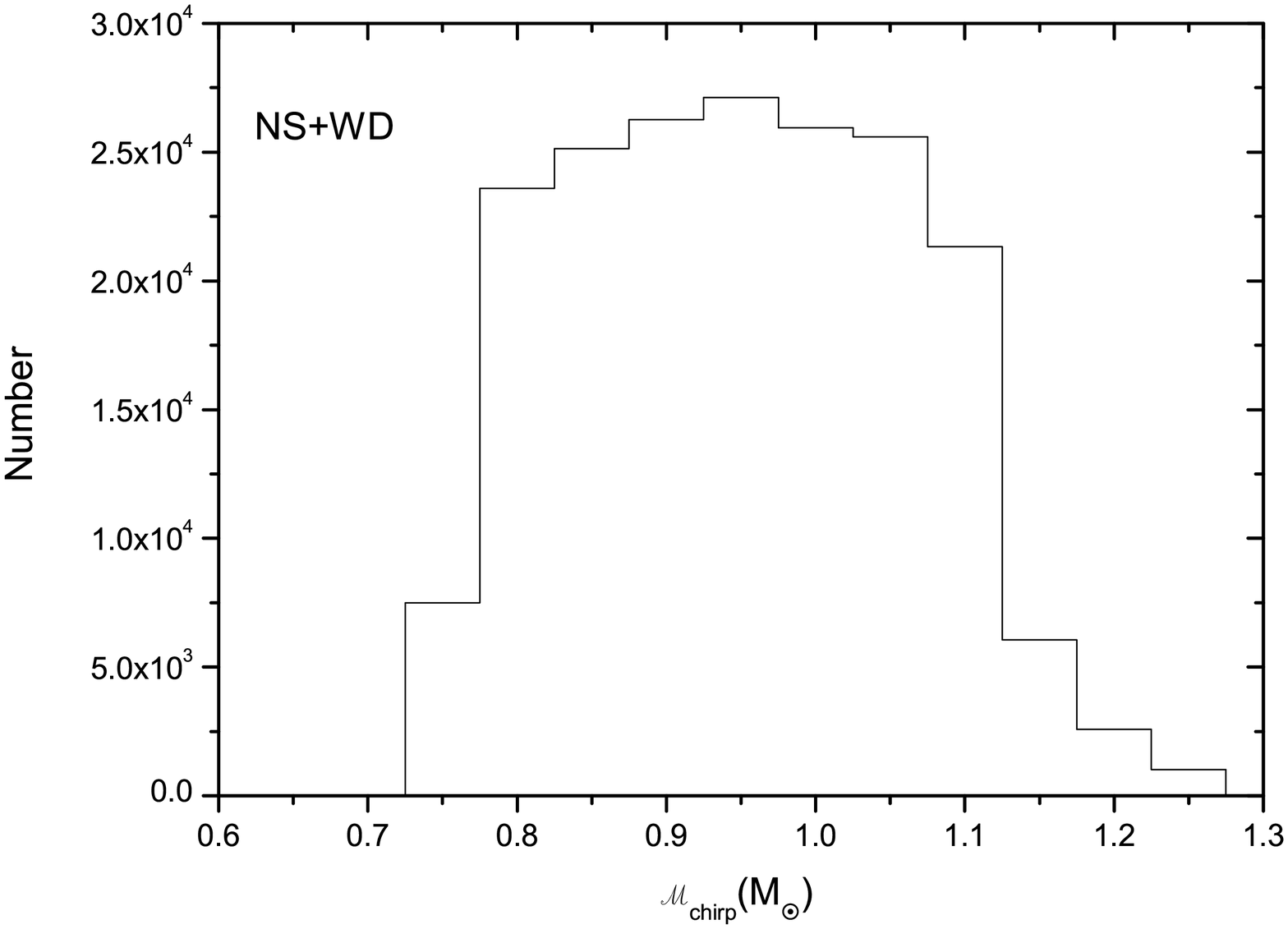}
\includegraphics[bb=80 100 550 600,height=2.2in,width=2.3in]{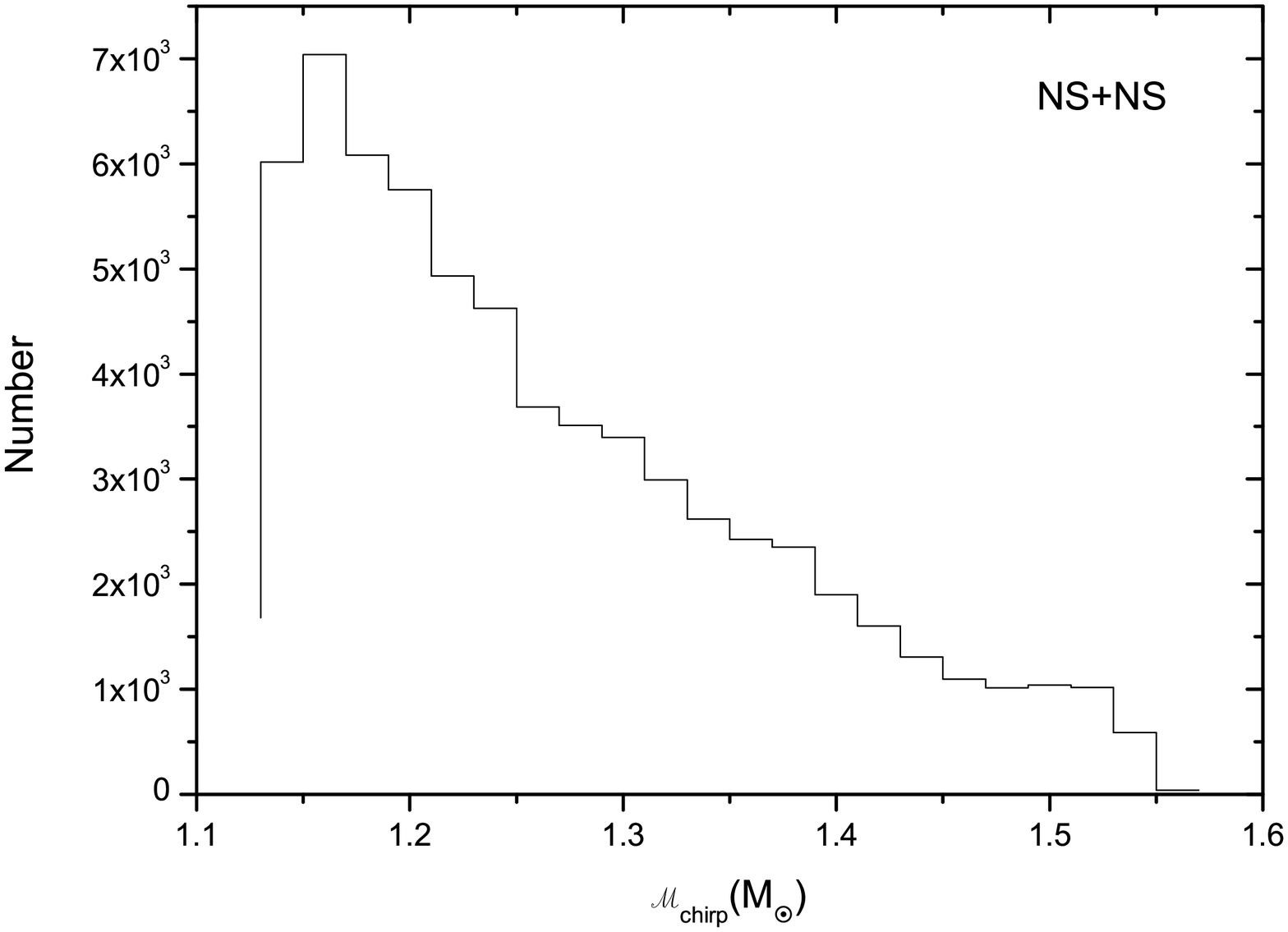}
\includegraphics[bb=250 50 800 600,height=2.2in,width=2.3in]{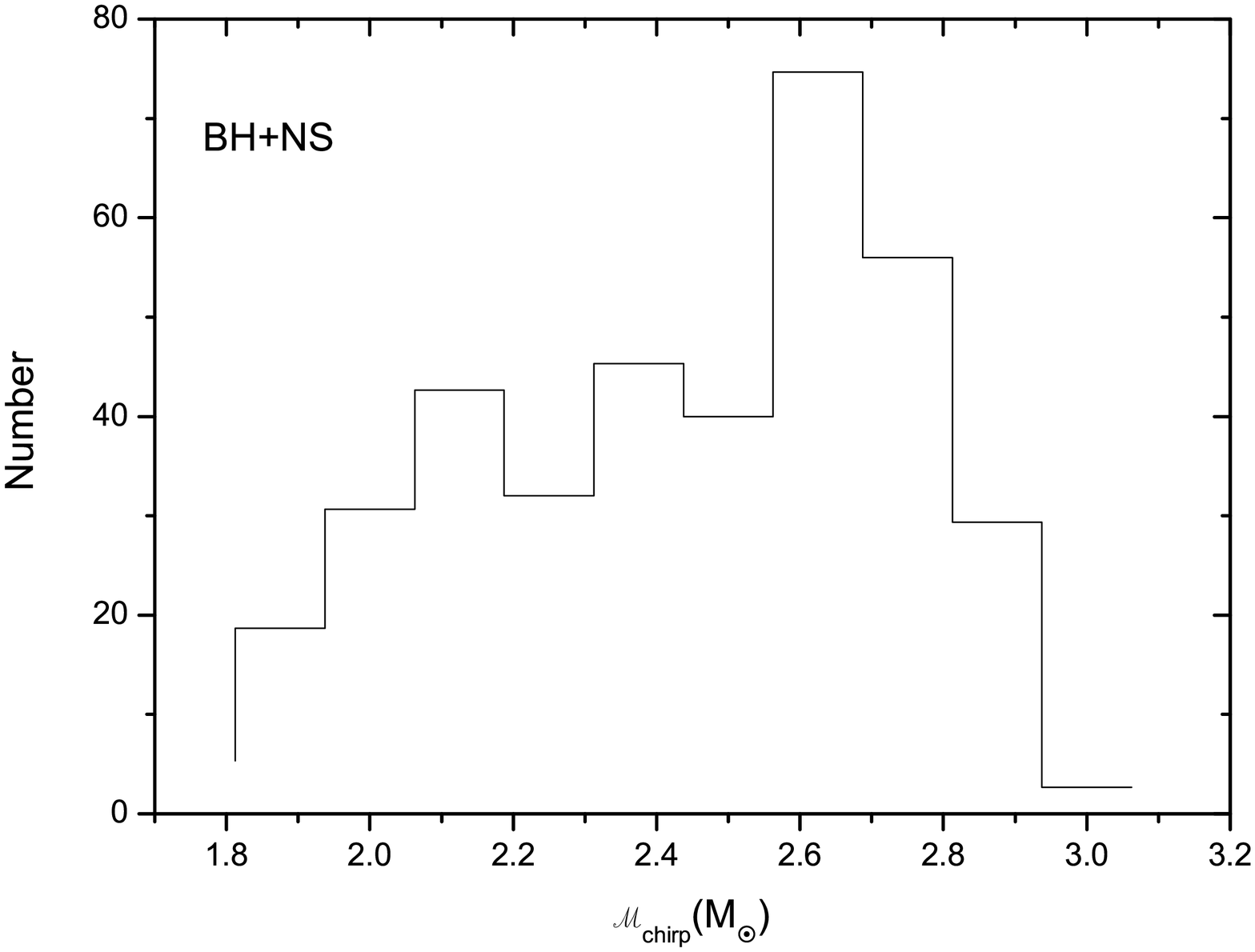}
\includegraphics[bb=80 50 550 600,height=2.2in,width=2.3in]{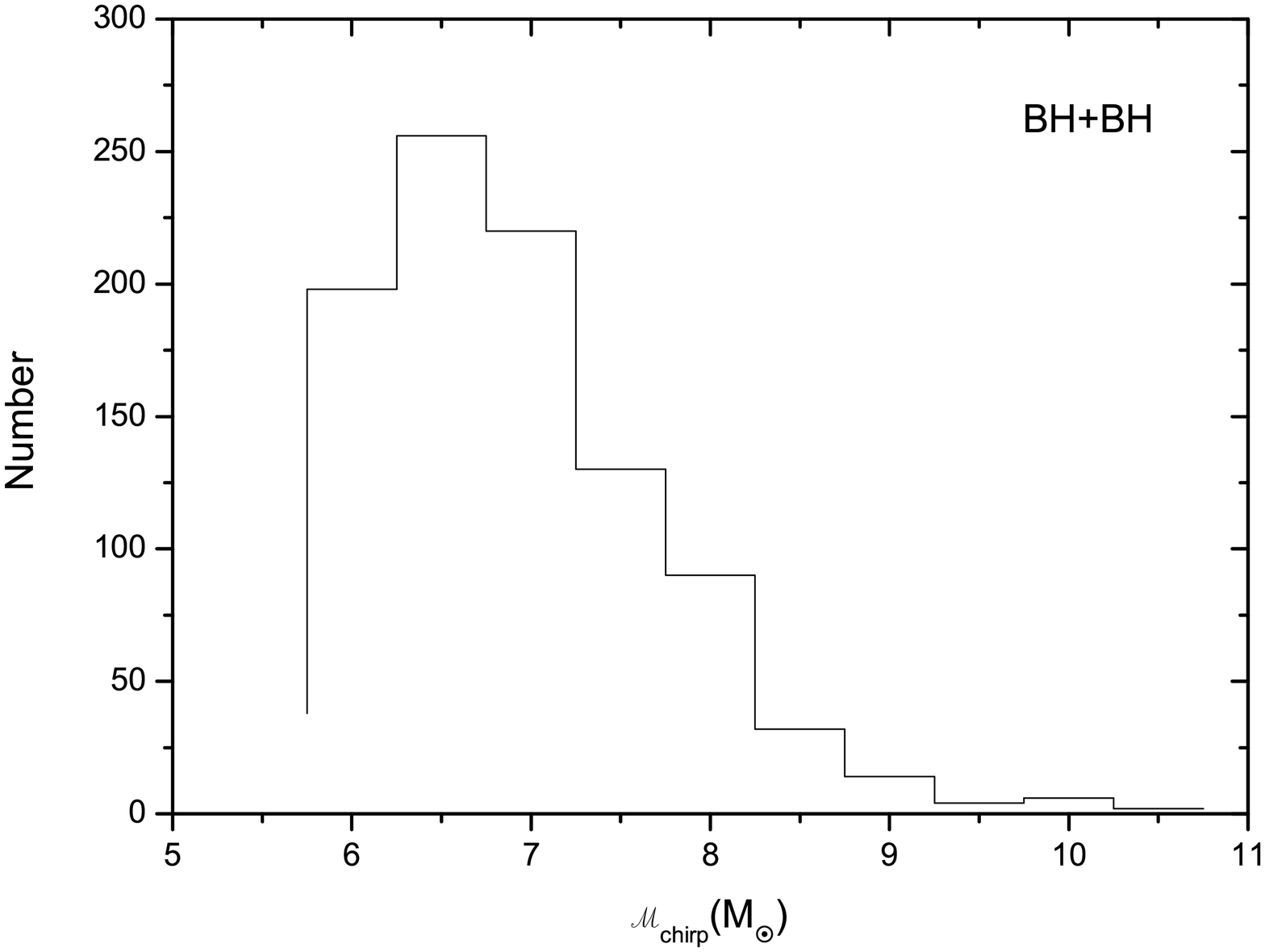}
\caption{The same as Figure 1 but for the distribution of chirp
masses for the DCO systems.}
\end{figure*}

\subsection{The chirp mass distribution of DCO systems}

A binary system radiates GWs and loses angular momentum, so the
separation decreases and the GW amplitude increases, which leads to
a measurable characteristic``chirp" signal. According to the
definition of ``chirp mass" for a binary system
 $\mathcal{M}_{\rm{chirp}}=m_{1}^{3/5}m_{2}^{3/5}(m_{1}+m_{2})^{-1/5}$, \citet{schu1996} displayed the frequency evolution
 of a binary due to GW radiation. In practice, for high-frequency binaries, this means that during a long observation time
 $T_{\rm obs}$, the ``chirp" signal can be detected by the eLISA detector \citep{evan1987}.

 In Figure 2, we present the chirp mass distribution for four types of DCOs in our model. From this we find that the averages
 of the chirp mass for the NS+WD, NS+NS, BH+NS, and BH+BH are 0.93$M_{\odot}$, 1.27$M_{\odot}$ , 2.50$M_{\odot}$, and 7.13$M_{\odot}$,
  respectively. Note that the chirp mass of PSR J0737-3039 (NS+NS) is 1.13$M_{\odot}$ \citep{lyne2004}, which is not much different
  from our calculations within a deviation $\sigma_{ \mathcal{M}_{\rm{chirp}}}=0.14 M_{\odot}$. We also find that the chirp mass
  distributions are very different for the four subclasses. For NS+WD systems, they bridge a narrow range: $ \mathcal{M}_{\rm{chirp}}\sim 0.72-1.27 M_{\odot}$
 and peak at $ \mathcal{M}_{\rm{chirp}}\sim 0.95 M_{\odot}$. For NS+NS systems, a well-known narrow peak occurs
 at $ \mathcal{M}_{\rm{chirp}}\sim 1.17 M_{\odot}$ with a stepping tail that extends to $ \mathcal{M}_{\rm{chirp}}\sim 1.57 M_{\odot}$.
 For BH+NS and BH+BH systems, the chirp mass distributions have a wide range: $ \mathcal{M}_{\rm{chirp}}\sim 1.8-3.2 M_{\odot}$ for BH+NS
 and $ \mathcal{M}_{\rm{chirp}}\sim 5.7-10.8 M_{\odot}$. A resolved chirping DCO binary can be used to measure the luminosity distance to the source directly,
  based on the chirp line for $T_{\rm obs}=2 $\,yrs
 (e.g., Equ.6 of Nelemans 2001) we find that the number of resolvable DCO binaries with the chirping signal is
 136.

 As displayed above, the significantly different results can be explained as follows. Firstly, a comparative
  distribution of NS+NS and BH+BH binaries with a similar shape depends not only on the choice of single star
  mass when BHs (or NS) are formed, but also, at some level, on the assumption
that an NS collapes to a BH when it accretes enough material to its
mass above 1.8 $M_{\odot}$ \citep{bomb1996}. This mass is not well
constrained. In this work, if an ONeWD or COWD accretes CO or ONe
material, which swelled up around these compact cores, and this new
mass exceeds the Chandrasekhar mass, $M_{\rm Ch}$, so the process of
electron capture on $^{24}$Mg nuclei leads to an AIC process
\citep{nomo1991} and the formation of an NS. Meanwhile, whether a
condition of thermonuclear
 explosion is actually not clear. The temperature produced at
 the core--disc boundary mostly depends on the accretion rate.
 If the temperature is hot enough to ignite carbon and oxygen,
 then the WD is converted to an ONeWD relies on competition
 between the rate of propagation of the flame inwards, which is judged by the opacity, and the cooling rate of the WD.
  All of the uncertainty can influence the mass determination of NS (or BH), and then affect
 the GW radiation characteristic parameter $ \mathcal{M}_{\rm{chirp}}$. Secondly, for BH+BH (or BH+NS) systems
 it is obvious that the BH mass is the decisive factor leading to the calculation of chirp mass. Note that the
 chirp mass coverage range ($ \mathcal{M}_{\rm{chirp}}\sim 1.8-10.5
M_{\rm \odot}$) predicted in the disk population of Galaxy is
consistent with the results of stellar mass population of Galactic
BHs with high metallicity environment \citep{ziol2010}. Thirdly, the
phase of dynamical mass transfer plays an important role in the
formation BH or NS. For example, dynamical mass transfer of helium
or CO on to a COWD or ONeWD causes the formation of a thick
accretion disk around the more massive WD ($M+\Delta M$), and the
coalescence is happened over a viscous time-scale. If the mass of
$M+\Delta M > M_{\rm Ch}$, it explodes as a possible type Ia SN
\citep{bran1998}, leaving no remnant. But a spherically symmetric
evolution model of \citet{saio1998} suggested that the AIC NS can be
also formed rather than type Ia SNe although the mass greater than
$M_{\rm Ch}$. Meanwhile the Eddington limit \citep{came1967} is
considered in the accreted material of WD, NS or BH.
\begin{figure*}%multicols\BB\B7\BE\B3\CF²\BB\C4ܸ\A1\B6\AF\A3\AC\BBᵼ\D6\C2ͼ\D0λ\F2\D5߱\ED\B8\F1\B6\AAʧ\A1\A3
%ֻ\C4ܵ\B1ǰλ\D6ã\A8[]\D6в\CE\CA\FD\B1\D8\D0\EB\CAǴ\F3дH\A3\A9\A3\AC\D2\F2\B4\CB\D0\E8Ҫ\CAֶ\AF\B5\F7λ\D6\C3
\centering
\includegraphics[bb=90 20 400 300,height=2.2in,width=2.3in]{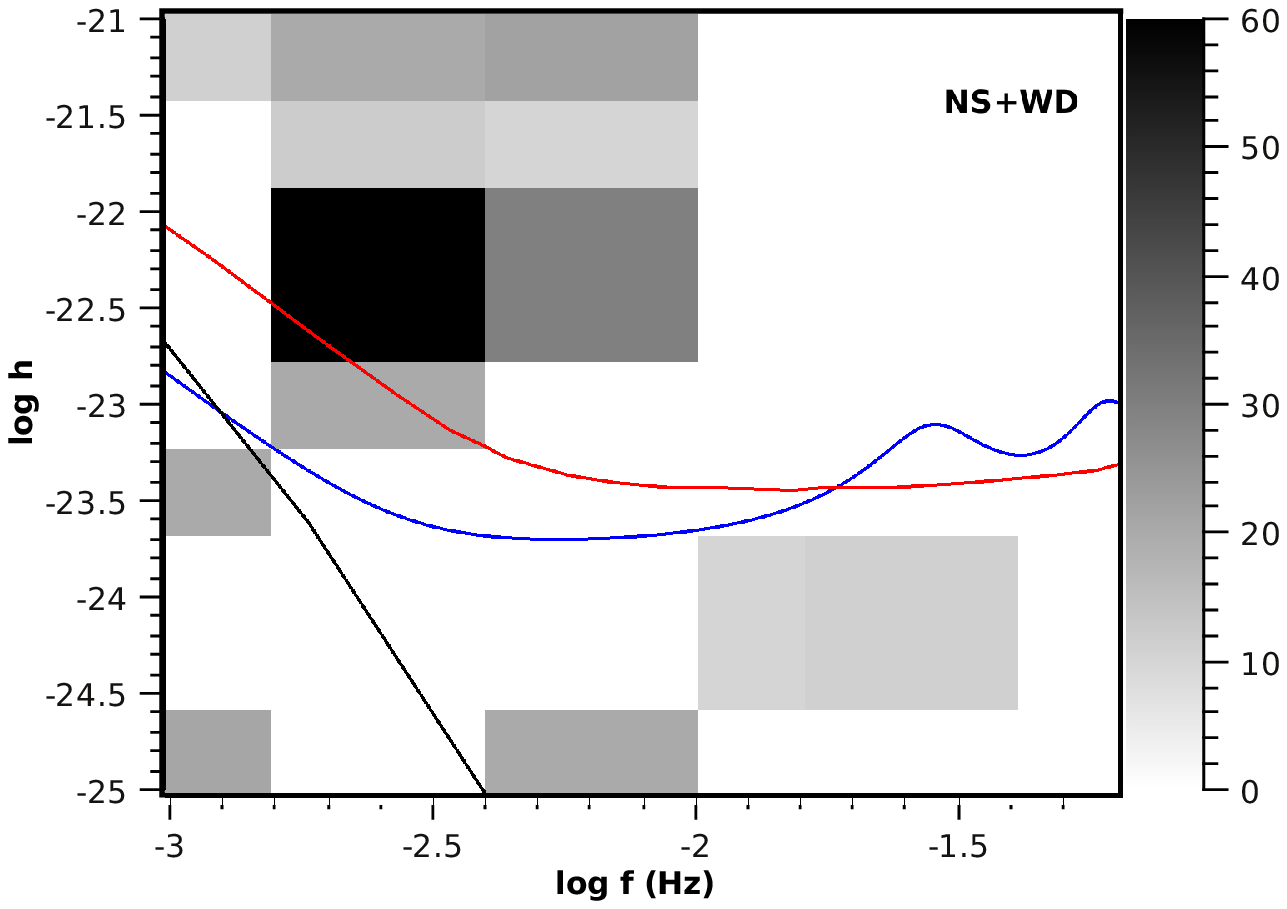}
\includegraphics[bb=10 20 320 300,height=2.2in,width=2.3in]{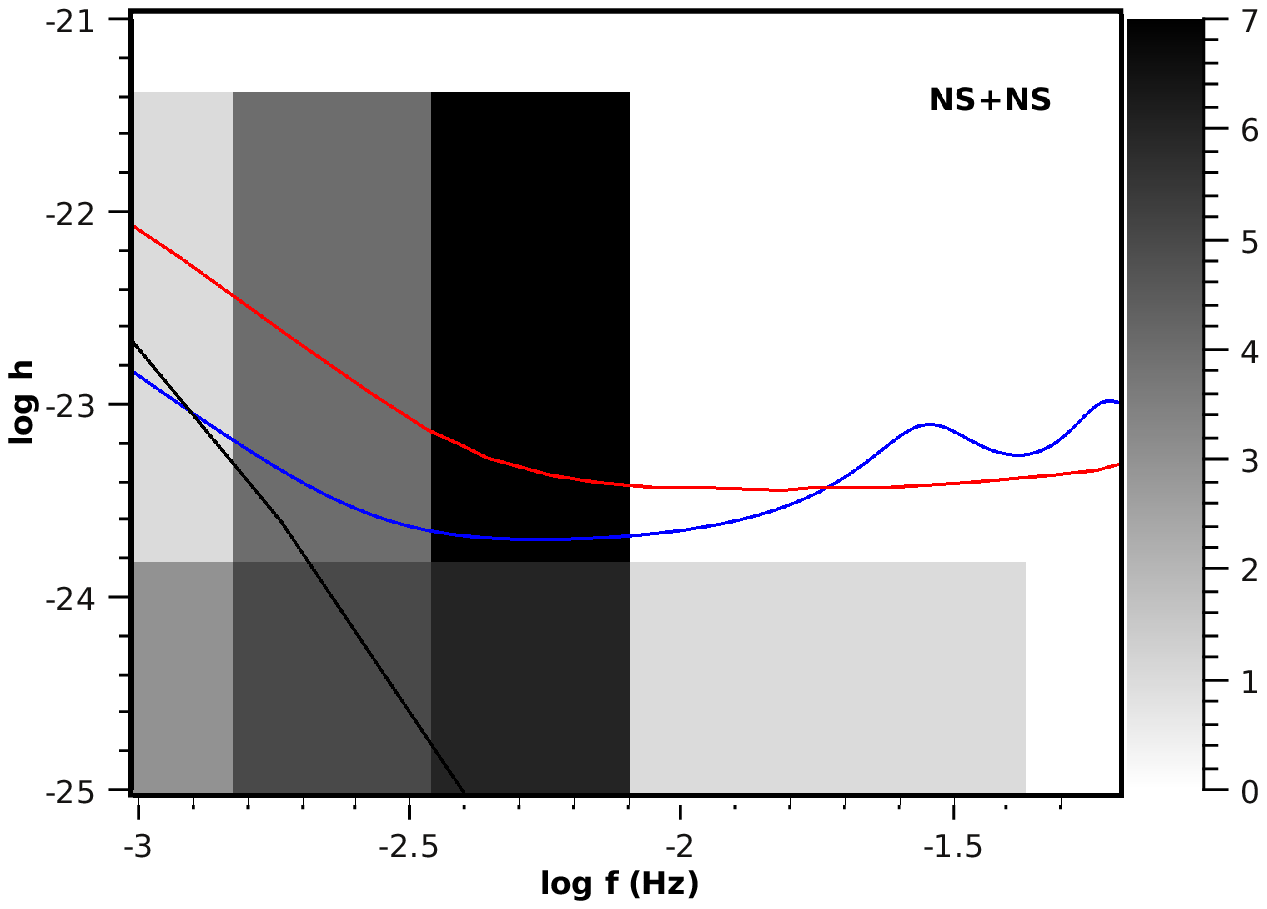}
\includegraphics[bb=90 20 400 300,height=2.2in,width=2.3in]{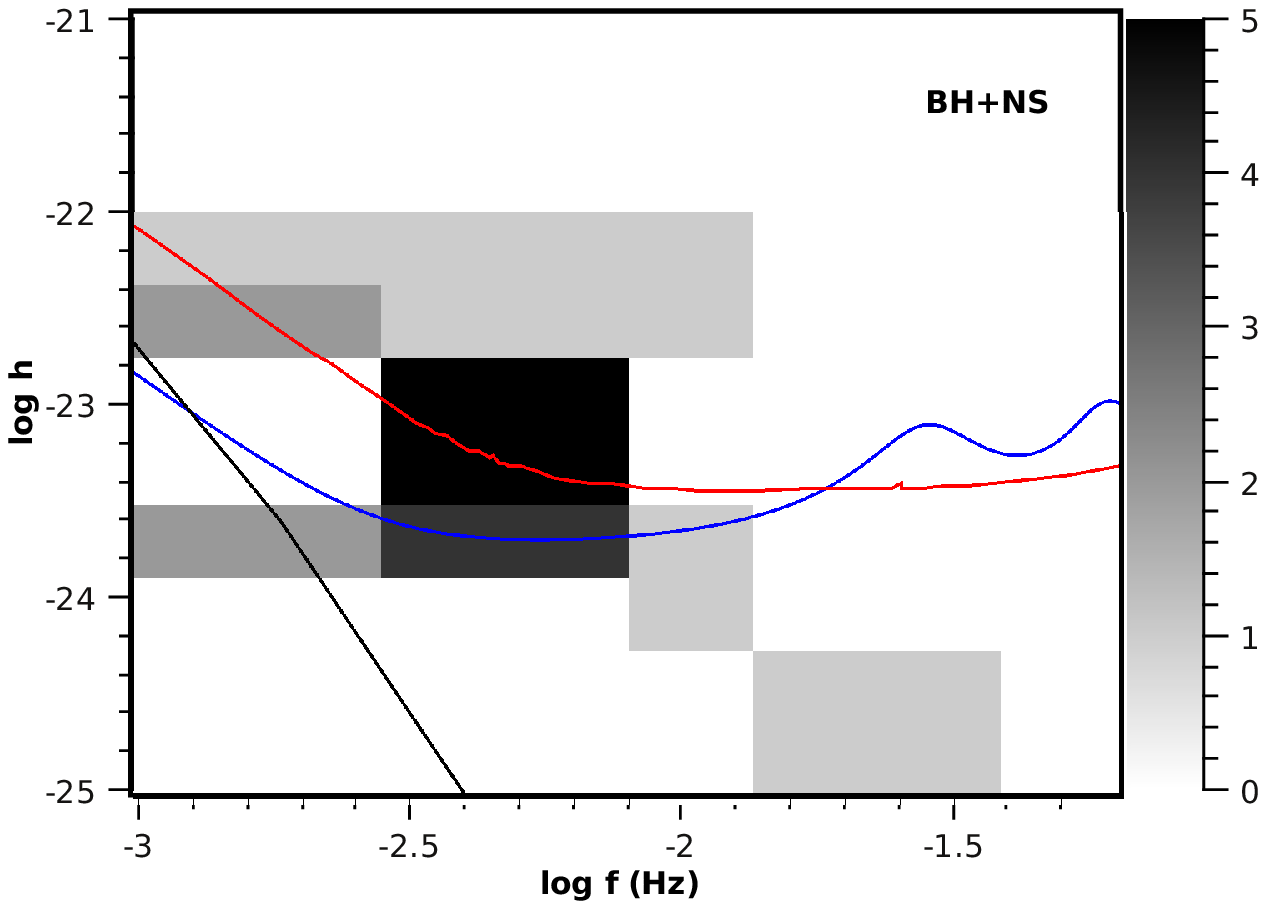}
\includegraphics[bb=10 20 320 300,height=2.2in,width=2.3in]{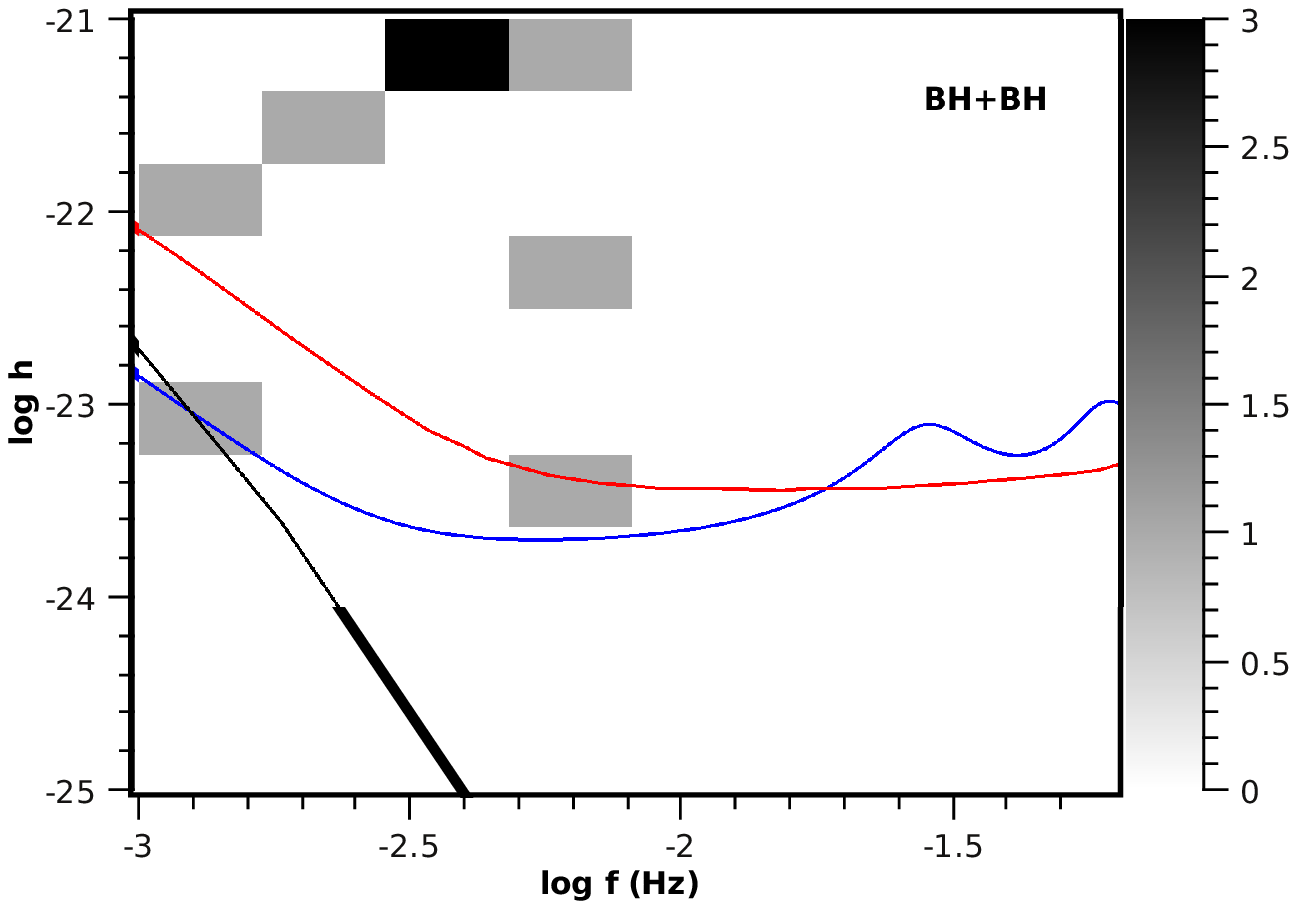}
\caption{Similar to Figure 1, but for the strain amplitude \emph{h}
as a function of the frequency for four types of DCO systems. In
each panel, the grey shade displays the density distribution of the
resolved systems; the black line gives the averaged GW foreground
due to a contribution of DWD systems in the Galaxy \citep{liu2009};
the red line stands for the expected eLISA sensitivity curve using
the simulator LISA--Code 2.0 \citep{petit08}. For a reference, we
also display
the sensitivity curve (blue line) of previous LISA detector \citep{lars2000}.} %ͼ\CC\E2
\label{fig:example2}%{}\D6С\B0fig:example2\A1\B1Ϊͼ\C3\FB\A3\AC\D2\FD\D3\C3ʱ\D3\C3\ref{fig:example2}
\end{figure*}

\subsection{The GW radiation signature for resolved sources}

According to the description of criterion in previous studies
\citep{liu2009, liu2010}, which can be summarized as ``one bin rule
+ the
 average foreground noise of DWDs + the eLISA sensitivity curve", the detectable numbers for four types
 of DCOs from the BPS simulation have been listed in Table 2. In Figure 3, we separate out the different
 kinds of DCOs in each panel. We plot the GW radiation distribution from the resolved sources (the grey shades)
 for four types of DCO systems. We also display the sensitivity limits of previous LISA \citep{lars2000}.
 The solid line stands for the average foreground that are produced by DWDs \citep{liu2009}. From this,
 in each panel we can see that only a few DCOs can be detected by the eLISA
 detector, which has become less sensitive than previous LISA.

The main purpose of this study is to find out how many resolved DCO
sources we can explore with an instrument like the eLISA detector
from the GW radiation.  For ``one bin rule", we assume that at least
one frequency in the eLISA data can be detected as GW sources. Here
we refer to "one bin rule", we aim to see whether the individual DCO
sources can be detected by the eLISA detector with one resolvable
bin ($\Delta f = 1.6\times 10^{-8} $Hz). The DCO binaries, which are
the only ones in their corresponding resolvable bins and their
strains are higher than the noise curve (including the sensitivity
curve of eLISA detector and the foreground noise foor of DWDs ), are
called detectable sources. A detected source with frequency \emph{f}
and strain amplitude \emph{h} that is observed over a time $T_{\rm
obs}$ will appear in the Fourier spectrum of the data as a single
spectral line. We note that this method is applicable to the
Galactic population of detached binaries (e.g., DWDs and DCOs), and
is discussed by the characterization GW signal in a
mass-transferring system \citep{timp2006}. For ``the average
foreground noise of DWDs",
 we use an exponential decay shape to modify the average foreground noise from detached DWDs in the Galaxy \citep{liu2009},
 which is not a Gaussian shape and does not include the loudest resolved WD populations.  For ``the eLISA sensitivity curve",
 comparing the GW strain amplitude from a DCO system with the average
 foreground noise of DWDs we can see that whether the individual sources can be detected by the eLISA detector.

Undoubtedly, comparing with previous studies we find that there
exist individual resolved DCO systems in the disk of the Galaxy
above the predicted average foreground noise of DWDs and the eLISA
sensitivity curve, which is likely to be used in the GW study. The
estimation of GW radiation agrees well with \citet{nele2001} and
\citet{belc2010}, although these authors used different binary
evolution assumptions in the BPS model for the underlying DCO
population. For example, the change of the binary orbital periods in
this work is governed by conservation of angular momentum rather
than energy, and during a CE evolution phase a DCO system can have a
shorter orbital period using the $\alpha$ formalism than ones using
$\gamma$ formalism. Therefore, for the detectable number of DCO
sources (e.g., NS+NS, BH+NS, BH+BH), \citet{nele2001} cited those
number as 38--124, 8--31 and 0--3, while \citet{belc2010} displayed
0.4--5, 0--0.6 and 0--3.5 respectively. Due to so many assumptions
as input as BPS simulations, we need to note that there exists
uncertainties in this prediction. We have studies the effect of
different physical parameters (e.g.,the ejection parameter of CE,
the mass transfer or stellar wind parameter, and tidal evolution)
and the various initial distributions of components (e.g, \emph{e},
IMF and \emph{q}) on the GW radiation from DCOs. We find that the
number of detectable sources of NS+WD, NS+NS, BH+NS ,and BH+BH can
range from 62 to 429, 7 to 52, 0 to 10, and 0 to 19, respectively.
Additionally, a form of understanding the CE evolution mechanism is
provided by the
 DCO binaries whose characteristics can be investigated if a significant amount of angular momentum and mass have been
 removed from the precursor system. Finally, we have not included any GW signals from EMRIs and super-massive black hole in-spirals in this work.

To summarize, we create a population of DCO systems using a Monte
Carlo simulation and discuss the importance of some physical process
and parameters on the formation of DCO systems in detail. Although
the presence of many uncertain factors (e.g., the CE evolution
question) should influence the outcome of
 this work, a simulation of the GW radiation from DCO systems has been still carried out using the BPS model.
 We present the distributions of orbital frequencies and chirp masses for DCO systems that are observable with the
  eLISA detector in a one-year observation. The total resolvable numbers of DCOs can take up to 200, and we find
  that the estimate for the number of resolved DCO sources ranges from 5 to 167.

\acknowledgments We thank Prof. Han Zhanwen at Yunnan Observatory
for valuable comments. This work is supported by the program of the
Xinjiang Natural Science Foundation (No. 2011211A104), Natural
Science Foundation (No. 11103054 and 11303080) and light in China's
Western Region (LCWR) (No. XBBS201022 and XBBS201221). This
project/publication was made possible through the support of a gran
from the John Templeton Foundation. The opinions expressed in this
publication are those of the authors and do not necessarily reflect
the view of the John Templeton Foundation. The funds from John
Templeton Foundation were awarded in a grant to The University of
Chicago which also managed the program in conjunction with National
Astronomical Observatories, Chinese Academy of Sciences (No.
100020101)

%% in parentheses, after the keyword, but they will not be verified.



\clearpage

%% the file, uncomment the \plotone, \plottwo, and
%% If you need a layout that cannot be achieved with \plotone or
%% \plottwo, you can invoke the graphicx package directly with the
%% \includegraphics command or use \plotfiddle. For more information,
%% please see the tutorial on "Using Electronic Art with AASTeX" in the
%% documentation section at the AASTeX Web site,
%% http://www.journals.uchicago.edu/AAS/AASTeX.
%%
%% The examples below also include sample markup for submission of
%% supplemental electronic materials. As always, be sure to check
%% the instructions to authors for the journal you are submitting to
%% for specific submissions guidelines as they vary from

%% This example uses \plotone to include an EPS file scaled to
%% 80% of its natural size with \epsscale. Its caption
%% has been written to indicate that additional figure parts will be
%% available in the electronic journal.

\clearpage

%% Here we use \plottwo to present two versions of the same figure,
%% one in black and white for print the other in RGB color
%% for online presentation. Note that the caption indicates
%% that a color version of the figure will be available online.
%%

%% This figure uses \includegraphics to scale and rotate the still frame

%% If you are not including electonic art with your submission, you may
%% mark up your captions using the \figcaption command. See the
%% No more than seven \figcaption commands are allowed per page,
%% so if you have more than seven captions, insert a \clearpage

%% Tables should be submitted one per page, so put a \clearpage before

%% Two options are available to the author for producing tables:  the
%% deluxetable environment provided by the AASTeX package or the LaTeX
%% table environment.  Use of deluxetable is preferred.
%%

%% Three table samples follow, two marked up in the deluxetable environment,
%% one marked up as a LaTeX table.

%% In this first example, note that the \tabletypesize{}
%% command has been used to reduce the font size of the table.
%% We also use the \rotate command to rotate the table to
%% landscape orientation since it is very wide even at the
%% Note also that the \label command needs to be placed

%% This table also includes a table comment indicating that the full
%% version will be available in machine-readable format in the electronic

\clearpage


%% The following command ends your manuscript. LaTeX will ignore any text

\end{document}